# New investigations on the transverse spin of structured optical fields

**Zhi-Yong Wang**[1*], **Bin Chen**[2], **Run-Xiang Wang**[1], **Shuang-Jin Shi**[1], **Qi Qiu**[1], **Xiao-Fei Li**[1]

[1]School of Optoelectronic Information, University of Electronic Science and Technology of China, Chengdu 610054, CHINA

[2]Shanghai Laboratory of HPSTAR, Center for High Pressure Science and Technology Advanced Research, Bldg 6, 1690 Cailun Rd, Pudong, Shanghai 201203, CHINA

*E-mail:   zywang@uestc.edu.cn

**Abstract**

Guided waves and surface waves can be taken as two typical examples of structured optical fields with the transverse spin. Analytical derivations are developed to demonstrate that (i) guided waves also carry the transverse spin that depends on the mean direction of propagation, which may have important applications in spin-dependent unidirectional optical interfaces; (ii) the quantization form of the transverse spin is for the first time revealed, which is not obvious and related to an ellipticity; (iii) from a unified point of view, the transverse spin can be attributed to the presence of an effective rest mass of structured optical fields; (iv) the transverse spin can also be described by the spin matrix of the photon field; (v) unlike a free optical field whose spin projection on the propagation direction is the only observable, owing to the effective rest mass, the spin projection of structured optical fields on other directions is also an observable, such that one can develop an optical analogy of spintronics. A preliminary idea about the potential applications of the transverse spin is presented, but an in-depth and complete study will be presented in our next work.

## 1. Introduction

As we know, light carries spin angular momentum associated with the polarization of light. As for a free optical field, only along its propagation direction the spin projection is an observable. However, as far as some structured optical fields are concerned, there is an intriguing *transverse* spin that was first described in 2012 by Bliokh and Nori [1], where the term "transverse" means that the spin is orthogonal to the mean momentum of the structured optical fields. Recently, there has been a rapidly growing interest in structured optical fields with the transverse spin [1-22], from which important applications in spin-dependent unidirectional optical interfaces have been found [9, 10, 13, 14, 23-32].

The purpose of the present paper is to provide a more rigorous and complete picture of the transverse spin, so as to pave the way for our future works based on the transverse spin. In view of the fact that guided waves inside a waveguide can be viewed as the superposition of two sets of TEM waves with the same amplitudes and frequencies, but reverse phases, to study the transverse spin of structured optical fields, we take guided waves (as propagating waves) and surface waves (as evanescent waves) as two typical examples. In contrast to Bliokh and Nori *et al*, we will describe the transverse SAM in the manner of being familiar to theoretical physicists as much as possible. Based on this paper, we will try to develop a new branch in optics (spinphotonics or spinoptics, say, it can be regarded as the optical analogy of spintronics) in our next work.

We work in the international system of units (SI), the four-dimensional (4D) metric tensor is taken as $\eta^{\mu\nu} = \text{diag}(-1,1,1,1)$ ($\mu,\nu = 0,1,2,3$), and let $x^\mu = (ct,\boldsymbol{r}) = (ct,x,y,z)$, $\partial_\mu = \partial/\partial x^\mu = (\partial/c\partial t, \nabla)$. Let $\varepsilon_0$ and $\mu_0$ denote the permittivity and permeability in vacuum, and then $c = 1/\sqrt{\varepsilon_0\mu_0}$ is the velocity of light in vacuum. Complex conjugation is denoted by $*$. Repeated indices must be summed according to the Einstein rule.

Throughout the paper we deal with structured optical fields (taking guided waves and planar surface waves as two examples) with the time-harmonic factor of $\exp(-i\omega t)$, and we consider only those materials (including vacuum) that are linear, isotropic and homogeneous. The electric field intensity $\boldsymbol{E}$ and the magnetic induction density $\boldsymbol{B}$ are determined from the potentials $A^{\mu} = (\varphi/c, \boldsymbol{A})$ or $C^{\mu} = (c\phi, \boldsymbol{C})$, and in our case, the gauge can be chosen such that $\varphi = 0$ and $\nabla \cdot \boldsymbol{A} = 0$, or $\phi = 0$ and $\nabla \cdot \boldsymbol{C} = 0$, where

$$\boldsymbol{E} = -\nabla\varphi - \partial \boldsymbol{A}/\partial t = -\partial \boldsymbol{A}/\partial t, \quad \boldsymbol{B} = \nabla \times \boldsymbol{A}, \tag{1}$$

$$\boldsymbol{E} = -\nabla \times \boldsymbol{C}, \quad \boldsymbol{B} = -\nabla\phi - \partial \boldsymbol{C}/c^2 \partial t = -\partial \boldsymbol{C}/c^2 \partial t. \tag{2}$$

When all field quantities are expressed as real functions, using Eqs. (1) and (2), one can prove that the total angular momentum of the electromagnetic field is given by [33-37]

$$\begin{aligned}
\boldsymbol{J} &= \int \varepsilon_0 \boldsymbol{r} \times (\boldsymbol{E} \times \boldsymbol{B}) \mathrm{d}^3 x \\
&= \varepsilon_0 \int E_i (\boldsymbol{r} \times \nabla) A_i \mathrm{d}^3 x + \varepsilon_0 \int \boldsymbol{E} \times \boldsymbol{A} \mathrm{d}^3 x, \\
&= \varepsilon_0 \int B_l (\boldsymbol{r} \times \nabla) C_l \mathrm{d}^3 x + \varepsilon_0 \int \boldsymbol{B} \times \boldsymbol{C} \mathrm{d}^3 x
\end{aligned} \tag{3}$$

provided the field falls off suitably as $|\boldsymbol{r}| \to +\infty$, where $\boldsymbol{r} = (x, y, z)$, $i, l = 1, 2, 3$ and repeated indices must be summed according to the Einstein rule. Then the total angular momentum can be decomposed into an orbital angular momentum depending on the reference frame and an angular momentum (spin) independent of the frame of reference. In particular, there are two alternative expressions of spin density, i.e., $\boldsymbol{s}^{\mathrm{e}} = \varepsilon_0 \boldsymbol{E} \times \boldsymbol{A}$ and $\boldsymbol{s}^{\mathrm{m}} = \varepsilon_0 \boldsymbol{B} \times \boldsymbol{C}$. In terms of complex field quantities, the time-averaged spin densities are

$$\boldsymbol{s}^{\mathrm{e}} = \varepsilon_0 \operatorname{Re}(\boldsymbol{E} \times \boldsymbol{A}^*)/2, \quad \boldsymbol{s}^{\mathrm{m}} = \varepsilon_0 \operatorname{Re}(\boldsymbol{B} \times \boldsymbol{C}^*)/2. \tag{4}$$

As for guided waves and surface waves with the time-harmonic factor of $\exp(-\mathrm{i}\omega t)$, Eqs. (1) and (2) imply that

$$\boldsymbol{A} = -\mathrm{i}\boldsymbol{E}/\omega, \quad \boldsymbol{C} = -\mathrm{i}c^2 \boldsymbol{B}/\omega. \tag{5}$$

In what follows, our work will be based on Eqs. (4) and (5).

## 2. Transverse spin of electromagnetic waves inside a waveguide

A systematical research on the transverse spin of electromagnetic waves inside a waveguide is still absent. Then, to present new investigations on the transverse spin of structured optical fields, we first take the transverse spin of the guided waves as an example.

**2.1 General considerations**

To present a self-contained argument, let us start from some fundamental materials in our notations and conventions. Let a Cartesian coordinate system be spanned by an orthonormal basis $\{e_1, e_2, e_3\}$ with $e_3 = e_1 \times e_2$, where the unit vectors $e_1$, $e_2$ and $e_3$ are respectively along the x-, y- and z-axes. A straight hollow-tube waveguide is placed along the z-axis, it has a rectangular cross section with perfectly conducting walls at $x = 0$, $x = a$, $y = 0$, and $y = b$ ($a \geq b$). The electromagnetic waves inside the waveguide are the sum of transverse magnetic (TM) and transverse electric (TE) fields [38, 39]. Let $k^\mu = (\omega/c, \boldsymbol{k})$ be the 4D wavenumber vector of photons inside the waveguide, where $\omega = c|\boldsymbol{k}|$, $k_1 = k_x = m\pi/a$ and $k_2 = k_y = n\pi/b$ ($m, n = 1, 2, 3...$ for the TM$_{mn}$ mode, $m = 1, 2, 3...$, $n = 0, 1, 2,...$ for the TE$_{mn}$ mode). The cutoff frequency of the waveguide is $\omega_{cmn} = c\pi\sqrt{(m/a)^2 + (n/b)^2}$. The electromagnetic waves inside the waveguide are, respectively,

$$\text{TM}_{mn}: \begin{cases} E_x = i\dfrac{m\pi}{a}\dfrac{k_z}{\omega_c^2}c^2 E_0 \cos(\dfrac{m\pi}{a}x)\sin(\dfrac{n\pi}{b}y)\exp[-i(\omega t - k_z z)] \\ E_y = i\dfrac{n\pi}{b}\dfrac{k_z}{\omega_c^2}c^2 E_0 \sin(\dfrac{m\pi}{a}x)\cos(\dfrac{n\pi}{b}y)\exp[-i(\omega t - k_z z)] \\ E_z = E_0 \sin(\dfrac{m\pi}{a}x)\sin(\dfrac{n\pi}{b}y)\exp[-i(\omega t - k_z z)], \; B_z = 0 \\ B_x = -i\dfrac{n\pi}{b}\dfrac{\omega}{\omega_c^2}E_0 \sin(\dfrac{m\pi}{a}x)\cos(\dfrac{n\pi}{b}y)\exp[-i(\omega t - k_z z)] \\ B_y = i\dfrac{m\pi}{a}\dfrac{\omega}{\omega_c^2}E_0 \cos(\dfrac{m\pi}{a}x)\sin(\dfrac{n\pi}{b}y)\exp[-i(\omega t - k_z z)] \end{cases} \quad (6)$$

$$\text{TE}_{mn}: \begin{cases} E_x = -\mathrm{i}\dfrac{n\pi}{b}\dfrac{\omega}{\omega_c^2}c^2 B_0 \cos(\dfrac{m\pi}{a}x)\sin(\dfrac{n\pi}{b}y)\exp[-\mathrm{i}(\omega t - k_z z)] \\ E_y = \mathrm{i}\dfrac{m\pi}{a}\dfrac{\omega}{\omega_c^2}c^2 B_0 \sin(\dfrac{m\pi}{a}x)\cos(\dfrac{n\pi}{b}y)\exp[-\mathrm{i}(\omega t - k_z z)] \\ E_z = 0, \ B_z = B_0 \cos(\dfrac{m\pi}{a}x)\cos(\dfrac{n\pi}{b}y)\exp[-\mathrm{i}(\omega t - k_z z)] \\ B_x = -\mathrm{i}\dfrac{m\pi}{a}\dfrac{k_z}{\omega_c^2}c^2 B_0 \sin(\dfrac{m\pi}{a}x)\cos(\dfrac{n\pi}{b}y)\exp[-\mathrm{i}(\omega t - k_z z)] \\ B_y = -\mathrm{i}\dfrac{n\pi}{b}\dfrac{k_z}{\omega_c^2}c^2 B_0 \cos(\dfrac{m\pi}{a}x)\sin(\dfrac{n\pi}{b}y)\exp[-\mathrm{i}(\omega t - k_z z)] \end{cases}, \quad (7)$$

where $E_0$ and $B_0$ are two real constants. For propagation modes $k_z$ is a real number, for evanescent modes $k_z = \mathrm{i}\beta$ with $\beta$ being a positive real number. Applying Eqs. (4)-(7), one can obtain ( $h = E_0$ for the TM mode, $h = cB_0$ for the TE mode, the same below)

$$\text{TM}_{mn}: \begin{cases} s_x^e = -\dfrac{n\pi}{b}\dfrac{k_z}{2\mu_0 \omega_c^2 \omega} h^2 \sin^2(\dfrac{m\pi}{a}x)\sin(\dfrac{2n\pi}{b}y) \\ s_y^e = \dfrac{m\pi}{a}\dfrac{k_z}{2\mu_0 \omega_c^2 \omega} h^2 \sin(\dfrac{2m\pi}{a}x)\sin^2(\dfrac{n\pi}{b}y) \\ s_z^e = 0 \end{cases}, \ s^m = 0. \quad (8)$$

$$\text{TE}_{mn}: \begin{cases} s_x^m = \dfrac{n\pi}{b}\dfrac{k_z}{2\mu_0 \omega_c^2 \omega} h^2 \cos^2(\dfrac{m\pi}{a}x)\sin(\dfrac{2n\pi}{b}y) \\ s_y^m = -\dfrac{m\pi}{a}\dfrac{k_z}{2\mu_0 \omega_c^2 \omega} h^2 \sin(\dfrac{2m\pi}{a}x)\cos^2(\dfrac{n\pi}{b}y) \\ s_z^m = 0 \end{cases}, \ s^e = 0. \quad (9)$$

Obviously, both Eq. (8) and Eq. (9) show that the spin angular momentum is transverse with respect to the mean momentum of the electromagnetic waves along the waveguide. BTW, one can show that $s^m = s^e = 0$ for all evanescent modes inside the waveguide.

Both Eqs. (8) and (9) show that, for each given position in the cross section of the waveguide, the sign of the transverse spin densities depends on that of $k_z$ (i.e., depends on the direction of propagation of the guided waves). Therefore, similar to the transverse spin in evanescent waves, the transverse spin of the guided waves will also have important applications in spin-dependent unidirectional optical interfaces.

To exhibit the density distribution of transverse-spin vectors within a cross section of the waveguide (i.e., at an arbitrary fixed $z$), we take the $TM_{11}$, $TM_{21}$, $TM_{22}$, $TE_{10}$, $TE_{11}$, $TE_{21}$, and $TE_{22}$ modes as examples, and show their spin-density distributions in Figs. (1a)-(1c) and Figs. (2a)-(2d), respectively, where the diagrammatic arrows represent local spin vectors, with the thickness of arrows being directly proportional to the spin density. For convenience let $a = b = 1$ and $\pi k_z h^2 / 2\mu_0 \omega_c^2 \omega = 1$ (note that what we care about is the relative distribution of the spin-density vectors).

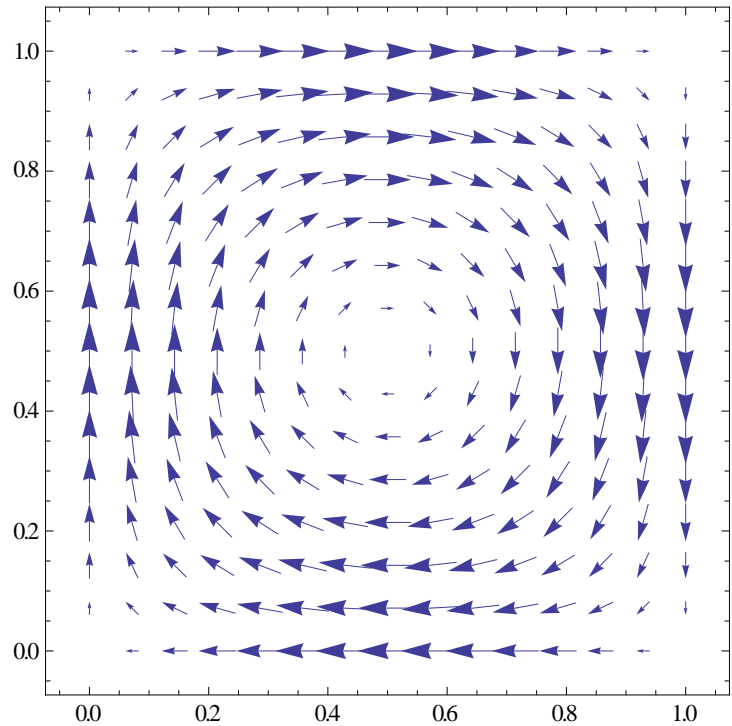

Fig. (1a) The transverse-spin density distribution of $TM_{11}$

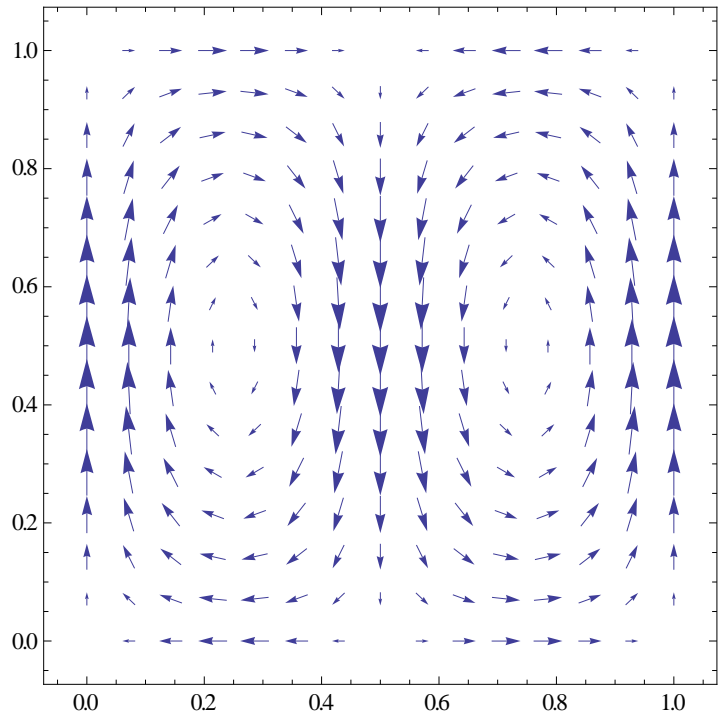

Fig. (1b) The transverse-spin density distribution of TM$_{21}$

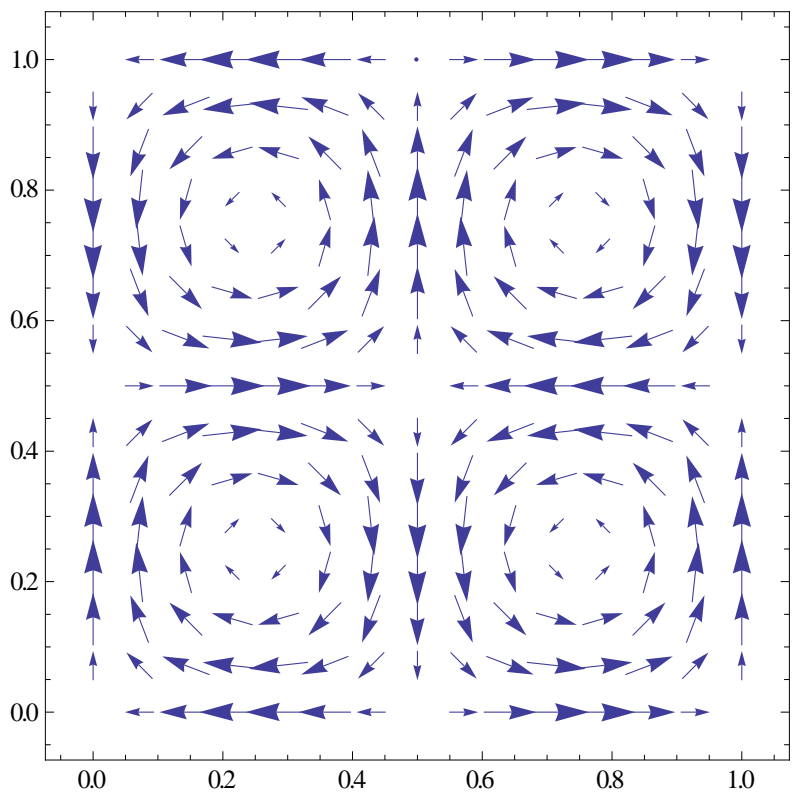

Fig. (1c) The transverse-spin density distribution of TM$_{22}$

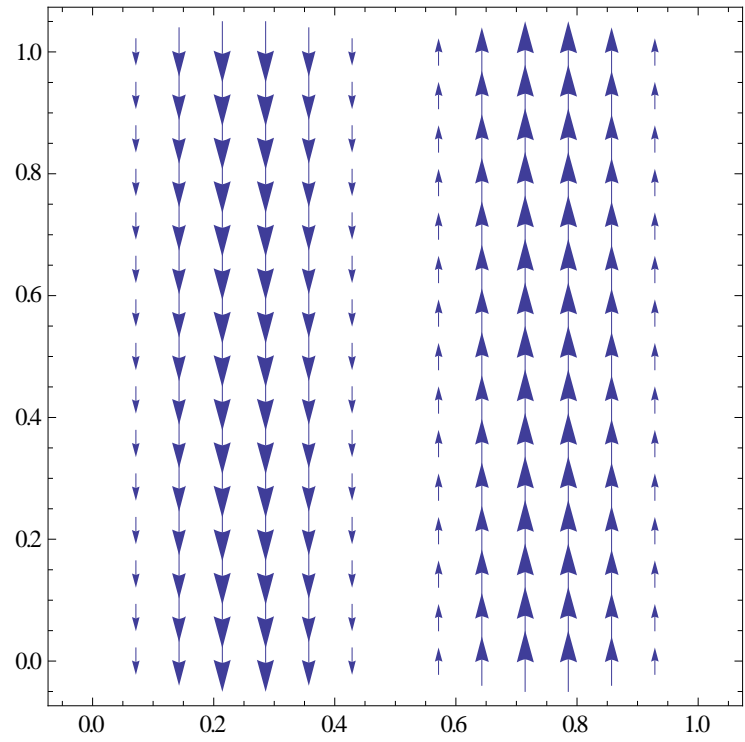

Fig. (2a) The transverse-spin density distribution of TE$_{10}$

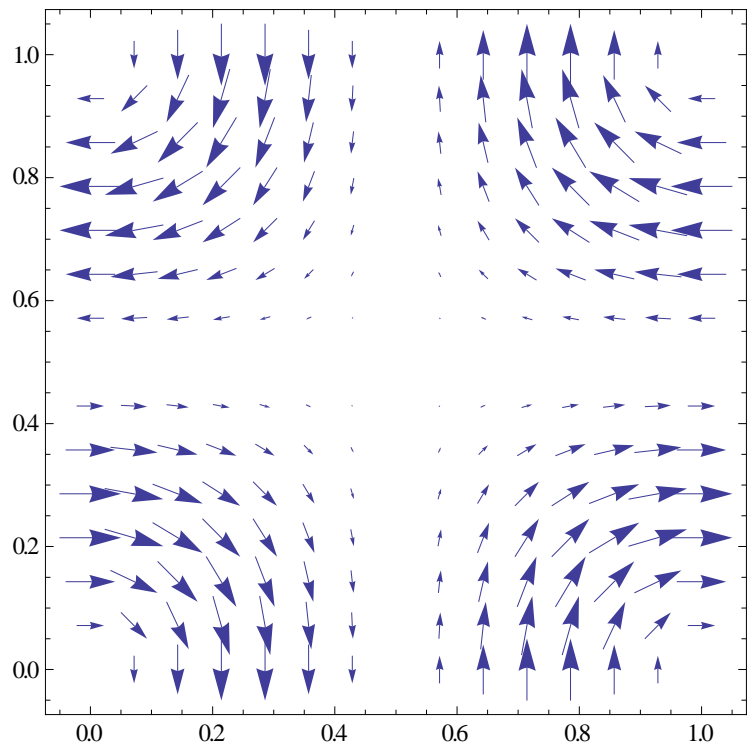

Fig. (2b) The transverse-spin density distribution of TE$_{11}$

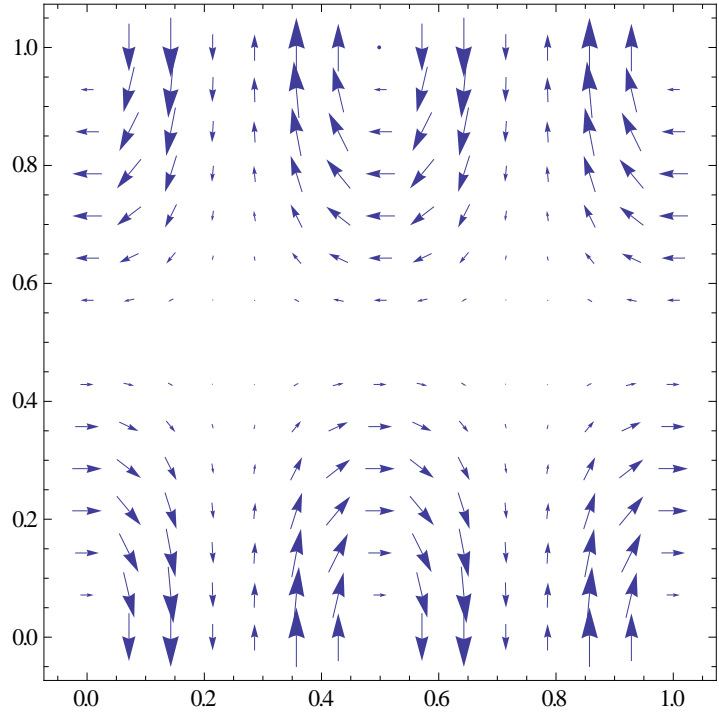

Fig. (2c) The transverse-spin density distribution of TE$_{21}$

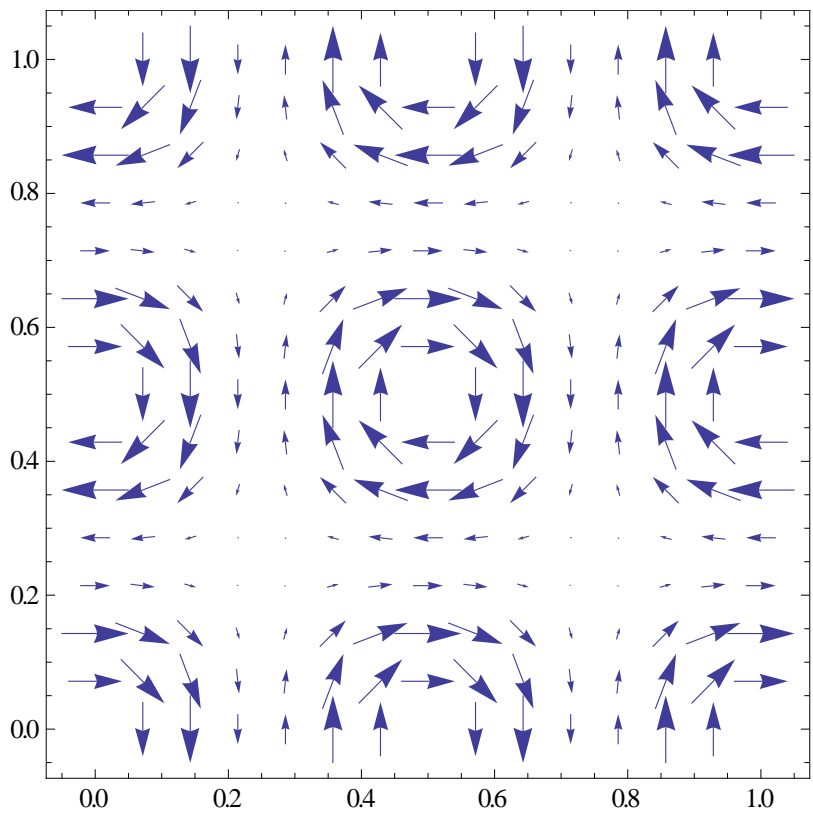

Fig. (2d) The transverse-spin density distribution of TE$_{22}$

**2.2 Some insights into the transverse spin of guided waves**

Up to now, people have not shown whether the transverse spin of structured optical fields is also quantized, and have focused their attention on the difference between the transverse and longitudinal spins. In particular, the previous descriptions of the transverse spin seem to imply that the transverse spin is not quantized. Therefore, we think it is valuable to reveal the quantization form of the transverse spin, which is actually far from an obvious or trivial fact. Let $L$ denote the length of the waveguide (and then $V = abL$ is the volume inside the waveguide), and define the total transverse spin $S_\perp$ of guided waves as

$$S_\perp^2 = \int_0^L \int_0^b \int_0^a (s_x^2 + s_y^2) \mathrm{d}x \mathrm{d}y \mathrm{d}z. \tag{10}$$

Substituting Eqs. (8) and (9) into Eq. (10) one has

$$S_\perp = \frac{\varepsilon_0 c k_z}{4\omega_c \omega} V h^2. \tag{11}$$

Note that $h = E_0$ for the TM mode and $h = cB_0$ for the TE mode. Eq. (11) shows that the total transverse spin $S_\perp$ also depends on the direction of propagation of the guided waves. On the other hand, the total time-averaged momentum of the guided waves is

$$\boldsymbol{P} = \int_0^L \int_0^a \int_0^b \frac{1}{2} \varepsilon_0 \mathrm{Re}(\boldsymbol{E} \times \boldsymbol{B}^*) \mathrm{d}x \mathrm{d}y \mathrm{d}z = (0, 0, P_z), \quad P_z = \frac{\varepsilon_0 \omega k_z}{8\omega_c^2} V h^2. \tag{12}$$

Likewise, using $\omega^2 = \omega_c^2 + c^2 k_z^2$ and $\omega_c = c\pi\sqrt{(m/a)^2 + (n/b)^2}$ one can show that the total time-averaged energy of the guided waves is

$$W = \int_0^L \int_0^a \int_0^b \frac{1}{4} \varepsilon_0 \mathrm{Re}(\boldsymbol{E} \cdot \boldsymbol{E}^* + c^2 \boldsymbol{B} \cdot \boldsymbol{B}^*) \mathrm{d}x \mathrm{d}y \mathrm{d}z = \frac{\varepsilon_0 \omega^2}{8\omega_c^2} V h^2. \tag{13}$$

Then the energy velocity along the waveguide is ( $c = 1/\sqrt{\varepsilon_0 \mu_0}$ )

$$\boldsymbol{v}_E = c^2 \boldsymbol{P}/W = (0,0,v), \quad v = P_z c^2/W = k_z c^2/\omega = \pm c\sqrt{1 - \omega_c^2/\omega^2}. \tag{14}$$

That is, the energy velocity equals to the group velocity of the guided waves. The quantization form of Eq. (13) can be expressed as follows (where the vacuum energy is discarded by taking the normal product of field operators):

$$W = \frac{\varepsilon_0 \omega^2}{8\omega_c^2} V h^2 = n\hbar\omega, \quad n = 1, 2, \ldots. \tag{15}$$

It follows from Eq. (15) that the quantization form of Eq. (12)

$$P_z = \frac{\varepsilon_0 \omega k_z}{8\omega_c^2} V h^2 = n\hbar k_z, \quad n = 1, 2, \ldots. \tag{16}$$

Comparing Eq. (15) or (16) with Eq. (11), and using Eq. (14), one can obtain,

$$S_\perp = 2\frac{ck_z \omega_c}{\omega^2} n\hbar = 2n\hbar \frac{v}{c}\sqrt{1 - v^2/c^2} = \pm n\hbar \sin 2\theta, \quad n = 1, 2, \ldots, \tag{17}$$

where $\cos\theta = |v/c| = |k_z c/\omega|$, $0 \leq \theta \leq \pi/2$. It should be emphasized that, though the guided waves with the transverse spin are linearly polarized within the polarization plane perpendicular to the mean propagation axis, they are elliptically polarized within a polarization plane containing the mean propagation axis, such that there is a factor of $\sin 2\theta$ on the right-hand side of Eq. (17), where $e = \tan\theta$ is the ellipticity of the polarization ellipse corresponding to the total transverse spin $S_\perp$. In fact, consider that the total quantities $S_\perp$, $P_z$, and $W$ are obtained via the 3D space integrals Eqs. (10), (12) and (13), respectively, let us define the average quantity over cross section of the waveguide:

$$\langle F \rangle = \frac{1}{ab} \int_0^a \int_0^b F \, dx \, dy. \tag{18}$$

Taking the TM mode as an example, one can show that the ellipticity of the polarization ellipse corresponding to the total transverse spin $S_\perp$ is

$$e = \tan\theta = h_L/h_\perp = |\omega_c/k_z c|, \quad 0 \leq \theta \leq \pi/2, \tag{19}$$

where

$$h_\perp = \sqrt{\langle E_x E_x^* + E_y E_y^* \rangle}, \quad h_L = \sqrt{\langle E_z E_z^* \rangle}. \tag{20}$$

Obviously, when $\theta = 0$ ($e = 0$) or $\theta = \pi/2$ ($e = \infty$), the elliptical polarization becomes a linear polarization, and then $S_\perp = 0$. Traditionally, the ellipticity of $e = \tan\theta$ is defined in the domain of $-\pi/4 \leq \theta \leq \pi/4$, but it is just a matter of convention.

Moreover, using Eq. (6) or (7) one has

$$\int_0^L \int_0^a \int_0^b \text{Re}(\boldsymbol{E} \cdot \boldsymbol{E}^* - c^2 \boldsymbol{B} \cdot \boldsymbol{B}^*) \mathrm{d}x \mathrm{d}y \mathrm{d}z = 0. \tag{21}$$

The factor of $\sin 2\theta$ on the right-hand side of Eq. (17) can be interpreted by means of Eqs. (8)-(10), (13) and (21). In particular, when the ellipticity $e = \tan \theta = 1$, Eq. (17) implies that

$$S_\perp = \pm n\hbar, \quad n = 1, 2, \dots. \tag{22}$$

That is, in the case of circular polarization (with the polarization plane containing the mean propagation axis), the quantization form of the total transverse spin is the same as that of the longitudinal spin. In a word, Eqs. (17) and (22) show that the transverse spin is also quantized.

**3. Effective rest mass of guided photons**

In the Cartesian coordinate system $\{\boldsymbol{e}_1, \boldsymbol{e}_2, \boldsymbol{e}_3\}$, we have $\omega = c|\boldsymbol{k}|$, $k_x = m\pi/a$, $k_y = n\pi/b$ ($m, n = 1, 2, 3\dots$ for the TM mode; $m = 1, 2, 3\dots$, $n = 0, 1, 2, \dots$ for the TE mode), $\omega_c = c\sqrt{k_x^2 + k_y^2}$, and then $\omega^2 = \omega_c^2 + c^2 k_z^2$. One can read $\boldsymbol{k} = \boldsymbol{k}_\perp + \boldsymbol{k}_L$, where

$$\boldsymbol{k}_\perp = k_x \boldsymbol{e}_1 + k_y \boldsymbol{e}_2, \quad \boldsymbol{k}_L = k_z \boldsymbol{e}_3, \tag{23}$$

stand for $\boldsymbol{k}$'s components perpendicular and parallel to the waveguide, respectively. According to the waveguide theory, the group velocity $\boldsymbol{v}_g$ (it is also the energy velocity) and phase velocity $\boldsymbol{v}_p$ of the guided waves along the waveguide are, respectively

$$\boldsymbol{v}_g = \boldsymbol{e}_3 c \sqrt{1 - (\omega_c/\omega)^2}, \quad \boldsymbol{v}_p = \boldsymbol{e}_3 c / \sqrt{1 - (\omega_c/\omega)^2}. \tag{24}$$

Then the guided waves satisfy all the de Broglie's relations of matter waves:

$$\epsilon^2 = \boldsymbol{p}^2 c^2 + m_0^2 c^4, \quad \epsilon = \hbar\omega, \quad \boldsymbol{p} = \hbar \boldsymbol{k}_L, \quad \boldsymbol{v}_g \cdot \boldsymbol{v}_p = c^2, \tag{25}$$

where $m_0 = \hbar\omega_c/c^2$ plays the role of rest mass, and then we call it the *effective rest mass* of single photons inside the waveguide. Using Eqs. (14) one has

$$\epsilon = m_0 c^2 \Big/ \sqrt{1 - v^2/c^2} \, . \tag{26}$$

This is exactly the relativistic energy formula. In fact, the velocity $v$ can be viewed as a relative velocity between an inertial observer and guided photons along the waveguide. On the other hand, in terms of the total energy and momentum of the guided waves, one can define the total effective rest mass $M_0$ of the guided waves as follows,

$$W^2 = \boldsymbol{P}^2 c^2 + M_0^2 c^4 = P_z^2 c^2 + M_0^2 c^4 \, . \tag{27}$$

It follows from Eqs. (12)-(14) that $W = M_0 c^2 \big/ \sqrt{1 - v^2/c^2}$. Note that $M_0$ is the total effective rest mass of the guided waves, while $m_0 = \hbar \omega_c / c^2$ is the effective rest mass of single photons in the guided waves. Using Eq. (25)-(27) and (15), one can obtain,

$$M_0 = n m_0 \, , \quad n = 1, 2, \ldots . \tag{28}$$

Eq. (28) implies that, along the direction of the waveguide, the relative velocity between single photons and the center of mass of the guided waves vanishes.

Now, let us consider an orthogonal decomposition for $p^\mu = (\hbar \omega / c, \hbar \boldsymbol{k})$ as follows:

$$p^\mu = p_T^\mu + p_L^\mu, \quad p_T^\mu \equiv (0, \hbar \boldsymbol{k}_\perp), \quad p_L^\mu = (\hbar \omega / c, \hbar \boldsymbol{k}_L) = (\epsilon / c, \boldsymbol{p}) \, . \tag{29}$$

Such an orthogonal decomposition is Lorentz invariant because of $p_{L\mu} p_T^\mu = 0$. Likewise, let $\boldsymbol{r}_L = z \boldsymbol{e}_3$ and $\boldsymbol{r}_\perp \equiv x \boldsymbol{e}_1 + y \boldsymbol{e}_2$, an orthogonal decomposition for $x^\mu = (ct, \boldsymbol{r})$ is

$$x^\mu = x_T^\mu + x_L^\mu, \quad x_T^\mu \equiv (0, \boldsymbol{r}_\perp), \quad x_L^\mu = (ct, \boldsymbol{r}_L) \, . \tag{30}$$

It is easy to show that

$$p_\mu x^\mu = p_{T\mu} x_T^\mu + p_{L\mu} x_L^\mu \, . \tag{31}$$

The operator $\hat{p}_\mu = i\hbar \partial_\mu = i\hbar \partial / \partial x^\mu$ represents the totally 4D momentum operator of photons inside the waveguide, while

$$\hat{p}_L^\mu = i\hbar \partial_{L\mu} = i\hbar \partial / \partial x_L^\mu \, , \tag{32}$$

represents the 4D momentum operator of photons moving along the waveguide. One can show that the electromagnetic field intensities (all denoted by $\psi$) inside the waveguide satisfy the following Klein-Gordon equation,

$$(\partial_{L\mu}\partial_L^\mu - m_0^2 c^2/\hbar^2)\psi = 0. \qquad (33)$$

It is worth notice that, under the Lorentz transformations consisting of Lorentz boosts and spatial rotations with respect to the direction of the waveguide (they form a Lorentz subgroup SO (1, 1)), Eqs. (27) and (33) are covariant and the effective rest mass $m_0 = \hbar\omega_c/c^2$ keeps invariant. In such a sense, the appearance of the effective rest mass of guided photons is related to the symmetry breaking from the Lorentz group SO (1, 3) to its subgroup SO (1, 1), which is similar to the fact that the emergence of a mass is always accompanied with some symmetry breaking.

The effective rest mass as the rest energy of photons inside the waveguide (i.e., the energy as the group velocity vanishes), arises by forming standing-waves along the cross-section of the waveguide. In other words, it arises by freezing out the degree of freedom of transverse motion, or, by localizing and confining the electromagnetic energy along the cross-section of the waveguide.

**4. New insights into the transverse spin of surface waves**

Let us consider planar surface waves with the time-harmonic factor of $\exp(-i\omega t)$. Consider a planar interface perpendicular to the x-axis separating two (linear, isotropic and homogeneous) media that are infinitely wide along the y-axis. Let $x=0$ be the boundary between the two semi-infinite half-spaces, where medium 2 in the space of $x>0$ is vacuum, and the refractive index of medium 1 in the space of $x<0$ is denoted as $\eta$. When an incident plane wave with the plane of incidence being the x-z plane approaches medium 2 from medium 1, it will form an electromagnetic mode (i.e., a surface wave) propagating along the interface in the z-direction but decaying exponentially in a direction pointing away from the interface into the space of $x>0$, provided that its angle of incidence $\phi$ satisfies $\eta\sin\phi>1$. Such a mode can be classified as (a) transverse electric (TE) or *s*-polarized, or (b) transverse magnetic (TM) or *p*-polarized, according to whether it possesses only a single electric or magnetic field component along the *y*-direction, $E_y$, $B_y$, respectively. For our purpose, let us only consider the surface wave in the space of $x>0$ (i.e. in vacuum near the interface). In our case, all nonvanishing components of the electromagnetic waves are [40], respectively

$$\text{TM mode:} \begin{cases} E_x = \frac{k_z}{\omega}c^2 a_0 \exp[i(k_z z - \omega t) - \kappa x] \\ E_z = \frac{-i\kappa}{\omega}c^2 a_0 \exp[i(k_z z - \omega t) - \kappa x], \\ B_y = a_0 \exp[i(k_z z - \omega t) - \kappa x] \end{cases} \quad (34)$$

$$\text{TE mode:} \begin{cases} E_y = b_0 \exp[i(k_z z - \omega t) - \kappa x] \\ B_x = -\frac{k_z}{\omega}b_0 \exp[i(k_z z - \omega t) - \kappa x], \\ B_z = \frac{i\kappa}{\omega}b_0 \exp[i(k_z z - \omega t) - \kappa x] \end{cases} \quad (35)$$

where $a_0$ and $b_0$ are two real constants, the decay constant $\kappa$ and the propagation constant $k_z$ satisfy, respectively,

$$\kappa = \frac{\omega}{c}\sqrt{\eta^2\sin^2\phi-1}, \quad k_z = \frac{\omega}{c}\eta\sin\phi, \quad k_z^2 c^2 = \kappa^2 c^2 + \omega^2. \quad (36)$$

As evanescent waves, the x-component of the wavenumber vector, $k_x = i\kappa$, is imaginary, provided that $\eta\sin\phi>1$, such that the z-component $k_z > \omega/c$. In light of the time-averaging theorem, the energy density of the surface waves is,

$$w = \frac{1}{4}\varepsilon_0 \text{Re}(\boldsymbol{E}\cdot\boldsymbol{E}^* + c^2\boldsymbol{B}\cdot\boldsymbol{B}^*) = \frac{k_z^2 c^2}{2\omega^2}\varepsilon_0 h'^2 \exp(-2\kappa x), \quad (37)$$

where $h' = ca_0$ for the TM mode and $h' = b_0$ for the TE mode (the same below). Likewise, the linear momentum density of the surface waves is,

$$p = \frac{1}{2}\varepsilon_0 \text{Re}(\bm{E} \times \bm{B}^*) = (0, 0, p_z), \quad p_z = \frac{k_z}{2\omega}\varepsilon_0 h'^2 \exp(-2\kappa x). \tag{38}$$

Then the energy velocity is

$$\bm{v}_E = c^2 \bm{p}/w = (0, 0, v), \quad v = \omega/k_z = \pm c\sqrt{1 - \kappa^2/k_z^2}. \tag{39}$$

Substituting Eqs. (5), (34) and (35) into Eq. (4), one can obtain that, for the TM mode, the time-averaged spin densities are

$$\bm{s}^e = (s_x^e, s_y^e, s_z^e) = (0, \varepsilon_0 h'^2 \frac{\kappa k_z c^2}{\omega^3}\exp(-2\kappa x), 0), \quad \bm{s}^m = (s_x^m, s_y^m, s_z^m) = (0, 0, 0). \tag{40}$$

Likewise, the time-averaged spin densities of the TE mode are

$$\bm{s}^e = (s_x^e, s_y^e, s_z^e) = (0, 0, 0), \quad \bm{s}^m = (s_x^m, s_y^m, s_z^m) = (0, \varepsilon_0 h'^2 \frac{\kappa k_z c^2}{\omega^3}\exp(-2\kappa x), 0). \tag{41}$$

Both Eq. (40) and Eq. (41) show that the transverse spin densities are transverse with respect to the mean momentum of the surface waves, and their sign depend on the sign of $k_z$ (i.e., depends on the direction of propagation of the surface waves).

To gain some more insights into the transverse spin densities given by Eqs. (40) and (41), let us consider the energy quantization of the surface waves (taking the normal product of field operators, one can drop out the vacuum energy):

$$W = \int_{-\infty}^{+\infty}\int_{-\infty}^{+\infty}\int_0^{+\infty} w \mathrm{d}x\mathrm{d}y\mathrm{d}z = \frac{k_z^2 c^2}{4\kappa\omega^2}\varepsilon_0 A h'^2 = n\hbar\omega, \quad n = 1, 2, ..., \tag{42}$$

where $A = \int_{-\infty}^{+\infty}\int_{-\infty}^{+\infty} \mathrm{d}y\mathrm{d}z$. Using Eqs. (38) and (39), one can obtain the quantized linear momentum of the surface waves:

$$P_z = \int_{-\infty}^{+\infty}\int_{-\infty}^{+\infty}\int_0^{+\infty} p_z \mathrm{d}x\mathrm{d}y\mathrm{d}z = \frac{k_z}{4\kappa\omega}\varepsilon_0 A h'^2 = \frac{v}{c^2} n\hbar\omega, \quad n = 1, 2, .... \tag{43}$$

The relation between $W$ and $P_z$ can be further exhibited via Eq. (50). Likewise, let $s_y = s_y^e, s_y^m$, using Eqs. (40) and (41) one can show that the total transverse spin is (note that $h' = ca_0$ for the TM mode and $h' = b_0$ for the TE mode)

$$S_y = \int_{-\infty}^{+\infty}\int_{-\infty}^{+\infty}\int_{0}^{+\infty} s_y \mathrm{d}x\mathrm{d}y\mathrm{d}z = \frac{k_z c^2}{2\omega^3}\varepsilon_0 A h'^2. \tag{44}$$

Eq. (44) implies that the sign of the total transverse spin depends on the direction of propagation of the surface waves. Using Eqs. (39) and (42) one has

$$S_y = 2n\hbar\kappa/k_z = 2n\hbar\tan\theta', \quad n=1,2,..., \tag{45}$$

where $\tan\theta' = \kappa/k_z$, $-\pi/4 \leq \theta' \leq \pi/4$. Though the surface waves with the transverse spin are linearly polarized within the polarization plane perpendicular to the mean propagation axis, they are elliptically polarized within a polarization plane containing the mean propagation axis, such that there is a factor of $e = \tan\theta'$ on the right-hand side of Eq. (45). In fact, $e = \tan\theta'$ is the ellipticity of the polarization ellipse corresponding to the total transverse spin $S_y$, i.e.,

$$e = \tan\theta' = \kappa/k_z = \begin{cases} E_z/E_x, & \text{for the TM mode} \\ B_z/B_x, & \text{for the TE mode} \end{cases}. \tag{46}$$

In contrary to the total transverse spin of the guided waves, the surface waves as evanescent waves, their total transverse spin is related to the ellipticity of the polarization ellipse in a different way. Moreover, in the case of evanescent waves, if we redefine the spin density as $s = (s^e + s^m)/2$, then one has $S_y = n\hbar\tan\theta'$ for the TM and TE modes. For the moment one has $S_y = n\hbar$ in the case of circular polarization.

## 5. Effective rest mass of surface waves

The surface waves studied above satisfy the dispersion relation of $\omega^2 = k_z^2 c^2 - \kappa^2 c^2$, it follows that their group velocity along the z-axis is larger than the velocity of light in vacuum, and the field quanta inside the surface waves seem to possess an imaginary rest mass. As a result, the field quanta inside the surface wave are thought of as owning tachyon properties [41, 42], and are called tachyon photons.

However, in contrary to the abovementioned guided waves with the group velocity being equal to the energy velocity, it is the energy velocity rather than the group velocity that represents the genuine motion velocity of the surface waves, and Eq. (39) shows that it is no longer a superluminal one. Therefore, it is advisable to define the effective rest mass density $\rho_0$ of the surface waves by means of the energy and momentum densities of the surface waves (given by Eqs. (37) and (38), respectively), that is, let

$$w^2 = \boldsymbol{p}^2 c^2 + \rho_0^2 c^4 = p_z^2 c^2 + \rho_0^2 c^4. \tag{47}$$

Using Eqs. (37), (38) and (47) one has

$$\rho_0 = \frac{w}{c^2}\sqrt{1-\frac{v^2}{c^2}} = \frac{\kappa k_z}{2\omega^2}\varepsilon_0 h'^2 \exp(-2\kappa x). \tag{48}$$

Eq. (48) implies that the field quanta inside evanescent waves should possess real mass (rather than unphysical imaginary mass). The total rest mass $M_s$ of the surface waves is:

$$M_s = \int_{-\infty}^{+\infty}\int_{-\infty}^{+\infty}\int_0^{+\infty} \rho_0 \mathrm{d}x\mathrm{d}y\mathrm{d}z = \frac{k_z}{4\omega^2}\varepsilon_0 A h'^2, \tag{49}$$

where $A = \int_{-\infty}^{+\infty}\int_{-\infty}^{+\infty}\mathrm{d}y\mathrm{d}z$. Using Eqs. (39), (42) and (43), one can prove that

$$W = \frac{M_s c^2}{\sqrt{1-v^2/c^2}}, \quad P_z = \frac{M_s v}{\sqrt{1-v^2/c^2}}. \tag{50}$$

It follows from Eqs. (42), (43) and (50) that, the single quanta of the surface waves have the energy of $\epsilon = \hbar\omega$, the momentum of $p = v\hbar\omega/c^2$, as well as the effective rest mass $m_s$ satisfying $M_s = nm_s$ ($n = 1, 2, ...$) and,

$$m_s = \frac{\hbar\omega}{c^2}\sqrt{1-v^2/c^2} = \frac{\hbar\kappa\omega}{c^2 k_z}. \tag{51}$$

Eq. (51) implies that,

$$\epsilon^2 = p^2 c^2 + m_s^2 c^4. \tag{52}$$

It is worth notice that, under the Lorentz transformations consisting of Lorentz boosts and spatial rotations with respect to the propagation direction of the surface waves (they form a Lorentz subgroup SO (1, 1)), Eq. (52) is covariant and the effective rest mass $m_s$ is invariant. In such a sense, the appearance of the effective rest mass is related to the symmetry breaking from the Lorentz group SO (1, 3) to the subgroup SO (1, 1).

**6. Transverse spin in terms of photonic spin matrix**

As we know, the spin of the photon field, i.e., the longitudinal spin in our context, can be described by means of spin operator or spin matrix. Here, we will for the first time study the transverse spin in terms of photonic spin matrix. To do so, let us consider the $(1,0) \oplus (0,1)$ representation of the group SL(2, C) that provides a six-component spinor equivalent to the electromagnetic field tensor [43, 44]. In what follows, the *column-matrix forms* of the vectors $\boldsymbol{E}$ and $\boldsymbol{B}$ are also denoted as $\boldsymbol{E}$ and $\boldsymbol{B}$, i.e., $\boldsymbol{E} = \begin{pmatrix} E_1 & E_2 & E_3 \end{pmatrix}^{\mathrm{T}}$, $\boldsymbol{B} = \begin{pmatrix} B_1 & B_2 & B_3 \end{pmatrix}^{\mathrm{T}}$ (the superscript T denotes the matrix transpose, the same below). In vacuum the $(1,0) \oplus (0,1)$ spinor can be expressed in two alternative forms [43, 44], i.e., the standard representation $\psi_S$ and the chiral representation $\psi_C$, where

$$\psi_S = \frac{1}{\sqrt{2}} \begin{pmatrix} \boldsymbol{E} \\ \mathrm{i}c\boldsymbol{B} \end{pmatrix}, \text{ or } \psi_C = \frac{1}{2} \begin{pmatrix} \boldsymbol{E} + \mathrm{i}c\boldsymbol{B} \\ \boldsymbol{E} - \mathrm{i}c\boldsymbol{B} \end{pmatrix}. \tag{53}$$

Under a Lorentz transformation parametrized by an antisymmetric tensor $\omega^{\mu\nu} = -\omega^{\nu\mu}$, $x^\mu \to x'^\mu = \Lambda^\mu{}_\nu x^\nu$, the $(1,0) \oplus (0,1)$ spinor $\psi$ transforms in the following manner:

$$\psi(x) \to \psi'(x') = \exp(-\mathrm{i}\omega^{\mu\nu} S_{\mu\nu}/2)\psi(x) = L(\Lambda)\psi(x), \tag{54}$$

where $L(\Lambda) = \exp(-\mathrm{i}\omega^{\mu\nu} S_{\mu\nu}/2)$ is the spinor representation of the Lorentz transformation $\Lambda$ (i.e., the $(1,0) \oplus (0,1)$ representation of SL(2, C)), and the antisymmetric tensor $S_{\mu\nu} = -S_{\nu\mu}$ is the 4D spin tensor of the field $\psi$, where

$$S_{lm} = \varepsilon_{lmn} \Sigma^n, \quad S_{0l} = -\mathrm{i}\alpha_l, \quad k,l,m = 1,2,3, \tag{55}$$

where $\varepsilon_{klm} = \varepsilon^{klm}$ denotes the totally antisymmetric tensor with $\varepsilon_{123} = 1$, and in the standard representation $\psi = \psi_S$, the matrices $\Sigma_l$ and $\alpha_l$ satisfy, respectively,

$$\Sigma = (\Sigma_1, \Sigma_2, \Sigma_3) = \begin{pmatrix} \tau & 0 \\ 0 & \tau \end{pmatrix}, \quad \alpha = (\alpha_1, \alpha_2, \alpha_3) = \begin{pmatrix} 0 & \tau \\ \tau & 0 \end{pmatrix}, \tag{56}$$

where the matrix vector $\tau = (\tau_1, \tau_2, \tau_3)$ consists of three components:

$$\tau_1 = \begin{pmatrix} 0 & 0 & 0 \\ 0 & 0 & -i \\ 0 & i & 0 \end{pmatrix}, \quad \tau_2 = \begin{pmatrix} 0 & 0 & i \\ 0 & 0 & 0 \\ -i & 0 & 0 \end{pmatrix}, \quad \tau_3 = \begin{pmatrix} 0 & -i & 0 \\ i & 0 & 0 \\ 0 & 0 & 0 \end{pmatrix}. \tag{57}$$

By means of the unitary transformations of $\psi_S \to \psi_C = U\psi_S$ and $S^{\mu\nu} \to S_C^{\mu\nu} = US^{\mu\nu}U^{-1}$, one can obtain the chiral representation from the standard representation, where the unitary matrix $U$ satisfies ($I_{n \times n}$ denotes the $n \times n$ unit matrix, $n = 2, 3, ...$)

$$U = U^\dagger = U^{-1} = \frac{1}{\sqrt{2}} \begin{pmatrix} I_{3\times 3} & I_{3\times 3} \\ I_{3\times 3} & -I_{3\times 3} \end{pmatrix}. \tag{58}$$

In fact, the chiral and standard representations of the $(1,0) \oplus (0,1)$ field are respectively analogous to the chiral and standard representations of the $(1/2, 0) \oplus (0, 1/2)$ field (the Dirac equation), but for the Pauli matrix vector $\sigma = (\sigma_1, \sigma_2, \sigma_3)$ being replaced with the matrix vector $\tau = (\tau_1, \tau_2, \tau_3)$. The matrix vector $\Sigma = I_{2\times 2} \otimes \tau$ satisfying $\Sigma \cdot \Sigma = 2$ is the 3D spin matrix of the photon field in the 6-component form given by Eq. (53). However, one can take $\tau$ as the 3D spin matrix vector of the photon field in the 3-component form (such as $E = (E_1 \ E_2 \ E_3)^T$, $B = (B_1 \ B_2 \ B_3)^T$). This is similar to the fact that $S = I_{2\times 2} \otimes (\sigma/2)$ satisfying $S \cdot S = 3/4$ is the 3D spin matrix of the Dirac field in the 4-component form, but $\sigma/2$ can be regarded as the 3D spin matrix of the Dirac field in the 2-component form.

For a beam of light propagating along an unit vector of $n = (n_1, n_2, n_3)$ with $n_1^2 + n_2^2 + n_3^2 = 1$, the spin projection operator along the propagation direction is $\tau \cdot n$, it satisfies the following eigenequation

$$(\boldsymbol{\tau} \cdot \boldsymbol{n})\boldsymbol{e}_\lambda(\boldsymbol{n}) = \lambda \boldsymbol{e}_\lambda(\boldsymbol{n}), \quad \lambda = \pm 1, 0. \tag{59}$$

The eigenvectors in Eq. (59) are, respectively,

$$\boldsymbol{e}_1(\boldsymbol{n}) = \boldsymbol{e}_{-1}^*(\boldsymbol{n}) = \frac{1}{\sqrt{2}} \begin{pmatrix} \dfrac{n_1 n_3 - i n_2}{n_1 - i n_2} \\ \dfrac{n_2 n_3 + i n_1}{n_1 - i n_2} \\ -(n_1 + i n_2) \end{pmatrix}, \quad \boldsymbol{e}_0(\boldsymbol{n}) = \boldsymbol{n} = \begin{pmatrix} n_1 \\ n_2 \\ n_3 \end{pmatrix}. \tag{60}$$

The eigenvector $\boldsymbol{e}_0(\boldsymbol{n})$ (parallel to the propagation direction) denotes the longitudinal polarization vector, while $\boldsymbol{e}_{\pm 1}(\boldsymbol{n})$ (perpendicular to the propagation direction) correspond to the right- and left-hand circular polarization vectors, respectively, and $\lambda = \pm 1, 0$ represent the spin projections in the direction of $\boldsymbol{n}$ (i.e., $\lambda = \pm 1, 0$ represent the helicities of photons). For a radiation field in free space, the $\lambda = 0$ photons corresponding to the longitudinal component describe the admixture of the longitudinal and scalar photons and exactly cancel each other [43]. However, for the guided and surface waves, the longitudinal direction is defined as being parallel to their mean propagation direction, such that the longitudinal component does exist. Or equivalently, owing to the presence of the effective rest mass, the $\lambda = 0$ photons can exist in the form of real photons with transverse spin, which we will further discuss later.

In view of the fact that an elliptical polarization can be described via a linear superposition of the circular polarization vectors $\boldsymbol{e}_1(\boldsymbol{n})$ and $\boldsymbol{e}_{-1}(\boldsymbol{n})$, for simplicity, in what follows we will only take into account the particular case of circular polarization. For the guided and surface waves, they propagate along the direction of $\boldsymbol{n} = (0, 0, 1)$, and then

$$\boldsymbol{e}_{\pm 1}(\boldsymbol{n}) = \frac{1}{\sqrt{2}} \begin{pmatrix} 1 \\ \pm i \\ 0 \end{pmatrix}, \quad \boldsymbol{e}_0(\boldsymbol{n}) = \begin{pmatrix} 0 \\ 0 \\ 1 \end{pmatrix}. \tag{61}$$

A detailed discussion about the relations between electromagnetic field intensities and the polarization vectors of $e_{\pm 1}(n)$ and $e_0(n)$, see Ref. [43]. Let $E = \begin{pmatrix} E_1 & E_2 & E_3 \end{pmatrix}^{\mathrm{T}}$, Eq. (61) implies that $E = E_\perp + E_L$, and (as mentioned above, just the case of circular polarization is considered)

$$E_\perp = \frac{1}{\sqrt{2}} \begin{pmatrix} 1 \\ \pm i \\ 0 \end{pmatrix} E_\perp, \quad E_L = \begin{pmatrix} 0 \\ 0 \\ 1 \end{pmatrix} E_3, \tag{62}$$

where $E_\perp = |E_\perp|$, $E_3 = 0$ for the TE mode, and $E_3 \neq 0$ for the TM mode. However, Eq. (62) implies that there is a phase difference of $\pm \pi/2$ between $E_1$ and $E_2$, while the transverse components $E_1$ and $E_2$ of the surface and guided waves are in phase (this is why the longitudinal spin of the surface and guided waves vanishes, actually in the sense of quantum-mechanical average). Therefore, only the longitudinal components of the surface and guided waves correspond to the eigenfunctions of Eq. (59) with $n = (0,0,1)$.

Now, let us consider the spin projections along the direction perpendicular to the mean propagation direction of the surface and guided waves, i.e., the spin projections along a transverse vector that can be expressed as the linear combination of two basis vectors $\eta_1 = (1,0,0)$ and $\eta_2 = (0,1,0)$. Then, we will just study the spin projections along the direction of $\eta_1 = (1,0,0)$ and $\eta_2 = (0,1,0)$, respectively. Let $\eta = \eta_1, \eta_2$, it follows from $(\tau \cdot \eta) e_\lambda(\eta) = \lambda e_\lambda(\eta)$ that $\lambda = \pm 1, 0$, and

$$e_{\pm 1}(\eta_1) = \frac{1}{\sqrt{2}} \begin{pmatrix} 0 \\ \pm i \\ -1 \end{pmatrix}, \quad e_0(\eta_1) = \begin{pmatrix} 1 \\ 0 \\ 0 \end{pmatrix}, \quad e_{\pm 1}(\eta_2) = \frac{1}{\sqrt{2}} \begin{pmatrix} 1 \\ 0 \\ \mp i \end{pmatrix}, \quad e_0(\eta_2) = \begin{pmatrix} 0 \\ 1 \\ 0 \end{pmatrix}. \tag{63}$$

Note that the guided and surface waves propagate along the direction of $n = (0,0,1)$. Eq. (63) implies that, the transverse spin described by the *longitudinal* circular polarization vectors of $e_{\pm 1}(\eta_1)$, is due to a phase difference of $\pm \pi/2$ between the transverse component $E_2$ and the longitudinal component $E_3$. Likewise, the transverse spin described by the longitudinal circular polarization vectors of $e_{\pm 1}(\boldsymbol{\eta}_2)$, is due to a phase difference of $\pm\pi/2$ between the transverse component $E_1$ and the longitudinal component $E_3$. More generally, when an electromagnetic wave possesses a longitudinal (or transverse) spin, it should be elliptically polarized with a polarization plane being orthogonal (or parallel) to the mean propagation direction of the wave.

Because the photon field is massless, its propagation direction in free space is a special one, only along such a special direction the spin projection of free photons is an observable. In contrary to which, as a massive spin-1/2 particle, the Dirac electron has a measurable spin projection along arbitrary given direction (determined by a magnetic field, for example), such that Spintronics has been developed during the last twenty years. Nevertheless, as shown above, when the photon field has an effective rest mass, its spin projection along other directions is also an observable. As a result, one can develop a new discipline that can be called Spinphotonics or Spinoptics, which is an optical analogy of Spintronics. In fact, based on effects of spin-symmetry breaking in nanoscale structures caused by spin-orbit interaction, a new branch in optics—Spinoptics in Plasmonics, has recently been developed [45].

**7. Preliminary idea about the potential applications of the transverse spin**

The purpose of the present paper is to provide a more rigorous and complete picture of the transverse spin, so as to pave the way for our next work. Here we just mention a preliminary idea about the potential applications of the transverse spin.

In fact, by means of the transverse spin of a surface plasmon wave, the authors of Ref. [10] have demonstrated a reciprocal effect of spin-orbit coupling when the direction of propagation of the surface plasmon wave determines a scattering direction of spin-carrying photons, where the spin-orbit coupling effect is an optical analogue of the spin injection in solid-state spintronic devices (inverse spin Hall effect), and may be important for optical information processing, quantum optical technology and topological surface metrology.

When a structured optical field with the transverse spin propagates along the z-axis, it should be an elliptically polarized wave with its polarization plane containing the z-axis. For example, as far as the guided and surface waves studied above are concerned, for the TM mode, $s^e \neq 0$ whereas $s^m = 0$, which is due to the fact that the electric field intensity $E$ is elliptically polarized while the magnetic induction $B$ is linearly polarized; for the TE mode, $s^e = 0$ whereas $s^m \neq 0$, which is due to the fact that $E$ is linearly polarized while $B$ is elliptically polarized. To illustrate the possibility of developing a new branch in optics based on the transverse spin, let us assume that a structured optical field (SOF) with the transverse spin propagates along the z-axis, $\boldsymbol{E}_{SOF} = (E_x, 0, E_z)$ for the TM mode, $\boldsymbol{B}_{SOF} = (0, B_y, B_z)$ for the TE mode, that is, the xz-coordinate plane is the elliptically polarized plane of $\boldsymbol{E}_{SOF}$, the yz-coordinate plane is the elliptically polarized plane of $\boldsymbol{B}_{SOF}$. Let us mention two examples as follows:

1). When the structured optical field (SOF) impinges on a nanoparticle, the latter will become polarized or magnetized. Assuming an isotropic polarizability $\alpha$ or an isotropic susceptibility $\zeta$ (up to a proportional constant) for the nanoparticle, the SOF-induced dipole *p* or magnetic moment *m* (density) is related to the impinging spin-carrying SOF as $\boldsymbol{p} = \alpha \boldsymbol{E}_{SOF}$ or $\boldsymbol{m} = \zeta \boldsymbol{B}_{SOF}$, therefore, describes an ellipse in the xz plane or in the yz plane, such as the dipole or magnetic moment components satisfy,

$$p_z/p_x = E_z/E_x \propto \eta_1 \exp(\pm i\pi), \text{ or } m_z/m_y = B_z/B_y \propto \eta_2 \exp(\pm i\pi). \tag{64}$$

where $\eta_1$ and $\eta_2$ are two real functions. In return, the SOF-induced dipole *p* or magnetic moment *m* will radiate an elliptically polarized wave. Starting from Eq. (64), one can study the transfer of the transverse spin between the structured optical field and the nanoparticle. In fact, starting from the first equation of Eq. (64), the authors of Ref. [10] have demonstrated the inverse spin Hall effect of photons.

2). Starting from the Dirac Hamiltonian for an electron (mass *m*, charge *e*) in an electromagnetic field with the gauge potential $A^{\mu} = (\Phi, \boldsymbol{A})$, one can obtain an angular magnetoelectric (AME) coupling Hamiltonian [46, 47]:

$$H_{\mathrm{AME}} = \frac{e^2 \hbar}{4m^2 c^2} \boldsymbol{\sigma} \cdot (\boldsymbol{E} \times \boldsymbol{A}), \qquad (65)$$

where $\boldsymbol{\sigma} = (\sigma_x, \sigma_y, \sigma_z)$ represents Pauli spin matrices, *V* is the crystal potential, $\boldsymbol{E} = \boldsymbol{E}_{\mathrm{ext}} - \nabla V/e$ with $\boldsymbol{E}_{\mathrm{ext}} = -\partial \boldsymbol{A}/\partial t - \nabla \Phi$. Obviously, Eq. (65) describes the coupling interaction between the spin of electrons and the one of photons. Starting from Eq. (65), one can study the transfer of the transverse spin between the structured optical field and electrons.

In addition to 1) and 2) mentioned above, we will provide a more key example in our next work, it is related to the splitting of energy levels between spin-up and spin-down states of photons. For all these examples an in-depth and complete study will be presented, with the aid of the $(1,0) \oplus (0,1)$ representation of SL(2, C).

**8. Conclusions and outlook**

In view of the fact that guided waves propagating along a hollow waveguide can be viewed as the superposition of two sets of TEM waves with the same amplitudes and frequencies, but reverse phases, to further study the transverse spin of structured optical fields systematically, one can take guided waves (as propagating waves) and surface waves (as evanescent waves) as two representative examples.

Apparently, there are several kinds of transverse spin, which differ strongly in their origins and physical properties. However, from a unified point of view, the transverse spin can be attributed to the presence of an effective rest mass of structured optical fields. More concretely, when an electromagnetic wave has a transverse spin, it should be an elliptically polarized wave with its polarization plane being parallel to the mean propagation direction of the wave. In particular, for the TM mode, $s^e \neq 0$ whereas $s^m = 0$, which is due to the fact that $\boldsymbol{E} = i\omega\boldsymbol{A}$ is elliptically polarized while $\boldsymbol{B} = i\omega\boldsymbol{C}/c^2$ is linearly polarized, or equivalently, the $A_\mu$ field corresponds to the spin eigenstate of the TM mode (note that $A^\mu = (0, \boldsymbol{A})$ and $\boldsymbol{E} = i\omega\boldsymbol{A}$); likewise, for the TE mode, $s^e = 0$ whereas $s^m \neq 0$, which is due to the fact that the $C_\mu$ field corresponds to the spin eigenstate of the TE mode (note that $C^\mu = (0, \boldsymbol{C})$ and $\boldsymbol{B} = i\omega\boldsymbol{C}/c^2$). From the point of view of group theory, the appearance of the effective rest mass of structured optical fields is related to the symmetry breaking from the Lorentz group SO (1, 3) to its subgroup SO (1, 1).

Up to now, people have focused their attention on the difference between the transverse and longitudinal spins, and have not shown whether the transverse spin is also quantized. In particular, the previous descriptions of the transverse spin seem to imply that the transverse spin is not quantized. We show that the transverse spin of structured optical fields is also quantized, which is actually far from an obvious or trivial fact. In the case of circular polarization (with the polarization plane containing the mean propagation axis), the quantization expression of the transverse spin is the same as the one of the longitudinal spin. Moreover, the transverse spin is also associated with the spin matrix of the photon field.

When the photon field has an effective rest mass, its spin projection along a transverse direction is also an observable (while for a photon field in free space, only along its propagation direction the spin projection is an observable). As a result, one can develop a new branch in optics that can be called spinoptics, which can be regarded as the optical analogy of spintronics. Interestingly, the sign of the guided waves' transverse spin also depends on the direction of propagation (similar to the transverse spin of evanescent waves), which have important applications in spin-dependent unidirectional optical interfaces, and the quantum spin Hall effect of light can be studied via guided waves.

Based on the present work, we will make some new contributions to spinoptics and try to combine it with spintronics in our next work. In a word, the issue about the transverse spin of light is new and very important. It has both theoretical and practical interests. For example, it associates with quantum spin Hall effect of light, it has potential application in future spinoptics, all-optical technologies, and quantum information, etc.

**Acknowledgments**

The author would like to thank Professor F. Nori for his helpful comments and suggestions. The work is supported by the National Natural Science Foundation of China (No. 61271030) and the National Natural Science Foundation of China (No. 61308041).